# Exciton-plasmon Coupling and Electromagnetically Induced Transparency in Monolayer Semiconductors Hybridized with Ag Nanoparticles

*Weijie Zhao, Shunfeng Wang, Bo Liu, Ivan Verzhbitskiy, Shisheng Li, Francesco Giustiniano, Daichi Kozawa, Kian Ping Loh, Kazunari Matsuda, Koichi Okamoto, Rupert F. Oulton, and Goki Eda*

Dr. W. Zhao, S. Wang, Dr. I. Verzhbitskiy, Dr. S. Li, Dr. F. Giustiniano, Prof. G. Eda
Department of Physics
Centre for Advanced 2D Materials and Graphene Research Centre
National University of Singapore
2 Science Drive 3, 117542 (Singapore)
E-mail: (g.eda@nus.edu.sg)

Dr. B. Liu, Prof. K. P. Loh, Prof. G. Eda
Department of Chemistry
Centre for Advanced 2D Materials and Graphene Research Centre
3 Science Drive 3, 117543 (Singapore)

Dr. D. Kozawa, Prof. K. Matsuda
Institute of Advanced Energy
Kyoto University
Uji, Kyoto 611-0011 (Japan)

Prof. K. Okamoto
Institute for Materials Chemistry and Engineering
Kyushu University
Motooka, Nishi-ku, Fukuoka 819-0395 (Japan)

Prof. R. F. Oulton
Blackett Laboratory
Imperial College London
Prince Consort Road, London SW7 2BZ (United Kingdom)

Metallic nanostructures support collective surface oscillations of the electrons gas, known as localized surface plasmons (LSP) or localized surface Plasmon resonance (LSPR), driven by incident radiation.[1-3] The LSP resonance frequency ($\omega_{LSP}$) of silver and gold nanoparticles can be effectively tuned by size, shape and surrounding dielectric medium.[1-3] At the resonance condition, LSPs efficiently confine incident photon energy into deep-subwavelength volumes and dramatically enhance local electromagnetic (EM) fields.[1-3] In turn, the enhanced EM field strongly modifies the light-matter interaction in molecules, quantum confined nanostructures, and other excitonic systems residing within the fringing field of the nanoparticle.[1-4]

Coupling between excitons and LSPs occurs via a coherent dipole-dipole interaction.[1,2] Depending on the lifetime of the LSP and the excitons, the interaction can be in the weak- or strong-coupling regimes. Strong de-coherence effects, such as the fast relaxation of LSPs on the femtosecond scale, often limit the interaction to the weak coupling regime.[1,2,5] In this regime, LSPs modify the absorption cross-section and, correspondingly, excitation rate of the exctionic system through a local field enhancement ($\gamma_{exc} \propto |E|^2$, where $\gamma_{exc}$ is the excitation rate and $E$ is the amplitude of electric field). Similarly, the spontaneous emission rate ($\gamma_{rad}$) is also enhanced due to an increase in local density of optical states, known as the Purcell effect.[1,2] Since radiative and dissipative loss in metal particles also introduces the possibility of emission quenching by dark modes[6] and lossy surface waves,[7] the overall enhancement in quantum yield ($\eta = \gamma_{rad}/(\gamma_{rad}+\gamma_{non-rad})$) is therefore determined by the competition of radiative and non-radiative ($\gamma_{non-rad}$) rates.[1,2] To achieve a large quantum yield enhancement, local field and radiative rate enhancements need to be optimized while minimizing non-radiative loss.[1,2]

In the strong coupling regime, a coherent coupling between LSPs and excitons overwhelms all losses and results in two new mixed states of light and matter separated energetically by a Rabi splitting that exhibits a characteristic anti-crossing behavior the exciton-LSP energy tuning.[5,8,9] In this regime, a new quasi-particle (plexciton) forms with distinct properties possessed by neither original particle. The behavior of plexcitons needs to be described in a quantum electrodynamics scenario. However, most concepts on plexcitons still remain theoretical predictions and experimental demonstrations have been largely limited to a few solid state material systems, including molecules,[8-10] quantum wells,[11] and quantum dots.[5] An intermediate regime, sometimes called Fano interference in the plasmonics field, is also expected when the interaction between excitons and LSPs is not sufficiently strong.[2,5,8,9,12] Notably, Fano interference manifests as transparency dips in the LSP spectrum that are reminiscent of a Rabi splitting; however, the interference effect tends to rapidly fade with increasing detuning.[2,5]

Two-dimensional (2D) semiconductors such as atomically thin films of transition metal dichalcogenides (TMDs) are emerging as a new platform to investigate and exploit excitons plasmon coupling.[13-15,16-18] Group 6 TMDs are well-known layered materials with the stoichiometry of $MX_2$, where M is a group 6 transition-metal (Mo, W) and X is a chalcogen (S and Se).[19] Single layer $MX_2$ are direct gap semiconductors with band gaps in the visible to near IR frequency range. Excitons in single layer $MX_2$ are strictly oriented in-plane[20], strongly bound,[21,22] and exhibit large oscillator strength.[14,21,23] The robust mechanical structure of 2D crystals allows ease of integration with photonic crystal cavities,[16,24,25,26] optical microcavities,[27] and plasmonic nanostructures,[15,16,17] offering exciting opportunities for engineering light-matter interaction and realizing novel photodetectors,[28] photovoltaics,[13,29] light-emitting diodes[30] and lasers.[24, 25]

Several recent studies[15,16-18] showed that the hybridizing monolayer $MoS_2$ results in various degrees of exciton-plasmon coupling and correspondingly enhanced light absorption and emission. While clear evidence of plasmon-enhancement has been demonstrated, the material is limited to $MoS_2$ and the extent of exciton-plasmon coupling remains elusive. In this communication, we report on plasmon-induced modification of light-matter interaction in 2D $MX_2$ ($MoS_2$, $WS_2$, $MoSe_2$ and $WSe_2$) layers hybridized with Ag nanoparticles. The hybrid structures were prepared by thermal evaporation of a metal thin film on top of monolayer $MX_2$ and subsequent annealing. The size and morphology of the metal nanoparticles were tuned through the thickness of the metal film, annealing temperature, and the surface condition of $MX_2$ sheets. We show that photoluminescence (PL) intensity enhancement is achieved via both optical field enhancement and spontaneous emission rate enhancement. Comparative studies of different $MX_2$ materials reveal that the emission quantum yield of the bare material determines the competition between enhancement and quenching effects. It was found that the quantum yield enhancement in originally bright materials such as $WS_2$ requires a dielectric spacer layer to minimize quenching effects. Furthermore, we report room temperature observation of a distinct electromagnetically induced transparency window that appears in the scattering spectra of $MoSe_2$-nanoparticle system, which indicates a coherent dipole-dipole interaction between exciton and LSP in the strong coupling regime. Strong exciton-plasmon coupling even at room temperature highlights the potential of this unique hybrid system for electro-optical modulation devices.

Monolayer $MX_2$ were obtained by mechanical exfoliation of bulk single crystals onto quartz and $SiO_2$/Si substrates. Bulk crystals (natural) of $MoS_2$ were obtained from SPI Supplies while $WS_2$, $MoSe_2$ and $WSe_2$ crystals were grown by chemical vapor transport (CVT) methods.[31,32] **Figure 1**a shows the optical and fluorescence images of monolayer $MoS_2$ sample on a $SiO_2$(90nm)/Si substrate. All exfoliated samples were annealed in vacuum ($10^{-5}$ mbar) at 150~200 °C to remove surface contaminations and other adsorbates. Subsequently, a thin film of silver (6 nm thick) was deposited on the samples under ultrahigh vacuum (~$8\times10^{-7}$ mbar) by thermal evaporation followed with mild annealing at 150 °C for 40 minutes under nitrogen atmosphere. Upon annealing, the silver adlayer agglomerated into disk-shaped nano-particles (NPs) on monolayer $MX_2$ surface (insert image of Figure 1b and S1), presumably due to surface diffusion, similar to previously reported observations.[33,34] As shown in Figures 1a and 1b, the optical appearance of MoS$_2$ changed dramatically after formation of silver NPs. The PL intensity measured under identical conditions was found to be significantly stronger for MoS$_2$ with Ag NPs as clearly seen in the fluorescence microscope images shown in Figure 1d. This enhancement is attributed to plasmon-induced enhancement of incident light absorption in addition to emission rate enhancement as will be discussed below.

In order to eliminate contributions from interference due the SiO$_2$/Si substrate, we focus the following discussions on the results from samples prepared on a transparent quartz substrate. **Figure 2**a and 2b show optical images for a monolayer MoS$_2$ sample on quartz without and with Ag nanoparticles, respectively. Disk-shaped Ag NPs of various sizes ranging from 20 to 120 nm with thickness of ~ 10 nm were found to be uniformly distributed on MoS$_2$ surface as shown in Figure 2c. (Note that the morphology of the particles on MoS$_2$ was similar on both SiO$_2$/Si and quartz substrates.) The particles were highly packed with inter-particle spacing of several tens of nanometers or less. Electric field distributions for different optical excitation energies were simulated using finite-difference time-domain (FDTD) simulations. The structure of experimentally obtained nano-disks (Figure 2c) was used as input parameters for the simulations. Figure 2d shows the simulated 2D mapping of EM fields. Strong enhancement in EM field up to 50 times was found to occur in "hotspots" that are located at the narrow gaps of particles where strong dipole-dipole interactions between adjacent LSPs are expected.[1-3] The average enhancement factor of EM fields over the simulated area was less than 10 since a large fraction of the surface is effectively inactive.

In another set of samples, we modified the sample preparation to obtain Ag NPs with significantly smaller sizes. For this, we washed the surface of exfoliated $MX_2$ sheets with acetone and isopropanol prior to metal deposition. The presence of trace amount of solvent residues significantly affected the surface migration of Ag, resulting in formation of semi-spherical particles with an average size of around 34 nm. The smaller particle size is most likely due to surface impurities acting as nucleation centers and pinning sites.[33] Morphology of the particles did not change with annealing (Figure 2g and Figure S1). We refer to samples with the large (57 nm average size) nano-disks and small (34 nm average size) semi-spheres as LNP (large NP) and SNP (small NP), respectively, in the following text. FDTD simulation of Ag SNP shows inter-particle hotspots with strong EM field enhancement but with a lower overall enhancement similar to the case of Ag LNP.

Experimental extinction spectra of monolayer $MoS_2$ with and without Ag NPs are shown in Figures 2i and 2j along with the FDTD-simulated spectra of LNP and SNP. The extinction spectra of the hybrid system consist of features that can be attributed to $MoS_2$ excitons and LSP of the NPs. A broad peak centered at ~700 nm observed for $MoS_2$-Ag LNP hybrid and a broad background peak at ~ 500 nm for $MoS_2$-Ag SNP hybrid can be attributed to LSPR of Ag NPs based on the agreement with our FDTD simulation results. The broad LSP resonance spanning the visible to near IR spectrum can be explained by the large distribution of particle sizes, suggesting that both excitation optical field and spontaneous emission rate enhancement can be responsible for the observed PL enhancement.

The observed PL intensity of emitters is determined by several factors such as collection efficiency of the spectroscopy setup, excitation rate ($\gamma_{exc}$) and quantum yield ($\eta$). Here, the collection efficiency is treated as a constant under identical experimental conditions. Because the in-plane oriented excitons of $MoS_2$[20] efficiently couple to the vertically incident laser whose polarization is also in-plane the excitation rate is considered to be linearly proportional to $|E|^2$. Consequently, the enhancement factor (*EF*) of PL is determined by:

$$EF = \left|\frac{E\prime}{E}\right|^2 \cdot \frac{\eta\prime}{\eta} \qquad (1)$$

where the symbols with and without the prime represent the corresponding parameters in the presence and absence of Ag NPs, respectively. The EM field enhancement and Purcell effect are strongly dependent on the LSPR conditions. In the following, we consider quantum yield modification via the Purcell effect and the EM field enhancement to study the origin of the observed PL enhancement.

According to Fermi's golden rule, the radiative rate is proportional to the local photonic density of states, that is $\gamma_{rad} \propto \rho(\omega)$.[1,2] Therefore, the spontaneous emission enhancement should follow approximately the extinction spectra of LSPs.[35,36] In **Figure 3**a, PL spectra of bare $MoS_2$ and $MoS_2$-LNP hybrid are shown along with their intensity ratio (grey plot). The emission spectrum is dramatically enhanced at the low energy side (above 670nm) for the hybrid system, revealing a wavelength-dependent enhancement spectrum, which agrees well with the coupled system's extinction spectrum. In contrast, the emission spectrum profile doesn't show significant change for $MoS_2$-SNP hybrid system only showing notable changes below 650nm (see Supporting Figure S6). In this system, however, local field enhancement is mainly responsible for PL enhancement due to the small overlap between LSPR and exciton emission peaks.

Time-resolved PL is an effective approach for quantitatively studying the spontaneous emission rate enhancement due to Ag NPs.[2,35] However, the short lifetime of $MX_2$ excitons (<100ps) owing to the dominance of fast non-radiative relaxation[37] as well as partial quenching by the metal NPs presents significant challenges in the measurement and interpretation of data. In order to eliminate the effects of EM field enhancement, we studied the emission enhancement in the excitation saturation regime. As shown in Figure 3b, the PL intensities of both the bare and hybrid samples exhibit a sub-linear increase as a function of laser power at a 633 nm excitation. While the full saturation regime was not explored in this experiment to avoid sample damage, the experimental data were instead fitted with the empirical equation for saturation: $I_{PL} = aP/(1 + bP)$, where $I_{PL}$ is PL intensity, $P$ is the excitation intensity, $a$ and $b$ are constants. By extrapolation, the PL *EF* in the saturation regime is determined to be about 6.8, which is effectively the enhancement of quantum yield, as under saturation, PL is independent of excitation intensity. The related radiative rate enhancement, i.e. Purcell factor for the radiative component, is also estimated to be around this value as will be discussed below.

In order to investigate the EM field enhancement, we studied the excitation wavelength dependence of PL *EF* for monolayer MoS$_2$ with Ag LNP and SNP as shown in Figure 3c and 3d, respectively. EM field enhancement factors were extracted from Raman scattering enhancement, which scales as $\beta \cdot |E'/E|^4$ where $\beta$ is a constant,[1,2,38] and plotted along with the PL *EF*. The obtained EM field enhancement factors are found to be relatively small (< 5) and only weakly dependent on excitation energy for both LNP and SNP hybrids in agreement with simulation results (See Supporting Information for simulation results, Figure S7). This was consistent with the simulated enhancement factors of the metal nanoparticle substrates. PL EF was consistently greater than EM field *EF* at all excitations for both hybrid systems. The observation is consistent with the contribution from quantum yield modification discussed above.

We note that deposition of Ag NPs on monolayer WS$_2$, which is significantly more luminescent than monolayer MoS$_2$ in the bare form,[32] consistently led to a reduction in the PL intensity shown in **Figure 4**a. Minor enhancement (~ 2) was achieved only by using a dielectric spacer layer such as hexagonal boron nitride (hBN) as shown in the insert of Figure 4b. This is attributed to the quenching component of the Purcell effect.[1, 2] We explain below that the striking contrast between the two materials originates from differences in their intrinsic quantum yield.

The quantum yield $\eta$ of photoluminescence is determined by ($\eta=\gamma_{rad}/(\gamma_{rad}+\gamma_{non-rad})$). In the presence of Ag NPs, the radiative emission rate is modified by Purcell effect and the expression becomes

$$\eta' = \frac{F_{rad}\gamma_{rad}}{F_{rad}\gamma_{rad}+\gamma_{non-rad}+F_{non-rad}\gamma_{rad}} \qquad (2)$$

where $F_{rad}$ is the partial Purcell factor leading to radiative decay and $F_{non-rad}$ is the remaining component of the Purcell Factor resulting in an effective non-radiative rate due to metal loss. (We assume that the intrinsic non-radiative rate does not change with deposition of Ag NPs.) The quantum yield enhancement then becomes

$$\frac{\eta'}{\eta} = \frac{F_{rad}}{\eta(F_{rad}+F_{non-rad}-1)+1} \qquad (3)$$

In the case of low quantum yield systems such as MoS$_2$, which typically exhibits $\eta = 10^{-3}$ ~ $10^{-4}$,[24] $\gamma_{non-rad}$ is orders of magnitude larger than $\gamma_{rad}$. For such systems, $\gamma_{non-rad} \gg F_{rad}\gamma_{rad} \gg \gamma_{rad}$ and the quantum yield enhancement becomes

$$\frac{\eta'}{\eta} = \frac{F_{rad}}{F_{non-rad}\cdot\eta+1} \qquad (4)$$

Equation (4) implies that materials with originally low quantum yield are only weakly affected by quenching effects of the metal. Indeed, in the limit of $F_{non-rad}\cdot\eta \ll 1$, the quantum yield enhancement is linearly proportional to the radiative Purcell factor for the spontaneous emission only. In the other limit of quantum yield approaching unity, quenching is unavoidable as can be seen from Equation (3). Thus, plasmon enhancement of moderately luminescent materials requires careful tuning of the ratio between $F_{rad}$ and $F_{non-rad}$ by controlling the spacing between the emitter and the metal.[1, 2] In Figure 4a, the relative quantum yields for monolayer $MoS_2$ and $WS_2$ with and without Ag NPs are compared. Monolayer $WS_2$, which is 100 times more luminescent than monolayer $MoS_2$, is significantly more prone to quenching when placed in contact with metal NPs.

We now return to the observation of a signature of strong coupling in the form of exciton-induced transparency dips of varying strengths in the various $MX_2$-NP samples. The two dips observed in the vicinity of the A and B exciton peaks for $MoS_2$-LNP system (Figure 2i) are an indication of Fano interference. Similar Fano-like features were recently reported for monolayer $MoS_2$ coupled with bow-tie-shaped plasmonic nanoantenna array at liquid nitrogen temperature.[17] The Fano interference arises from coherent dipole-dipole interaction between the discrete exciton states and a broad plasmon resonance state,[2, 5, 8, 9] as schematically shown in the inset of **Figure 5**. Our observation indicates that the coupling between exciton and plasmon enters the strong coupling regime even at room temperature and results in the large enhancement of spontaneous emission rate according to:[5]

$$F_{rad}\gamma_{rad} = \gamma_{rad} + 4\kappa^2/\Gamma_{LSP} \qquad (5)$$

where $\kappa$ is the exciton-plasmon coupling rate and $\Gamma_{LSP}$ is the linewidth of LSP. The coupling rate is dependent on the oscillator strength of exciton state and local field enhancement by LSP.[1,5] However, the high defect and doping concentration of monolayer $MoS_2$ (see Supplementary Figure S13) reduces oscillator strength of excitons[39,40] and is therefore detrimental to the coherent coupling.[40,41]

Monolayer $MoSe_2$ exhibits sharp exciton features (see Figure 5 and Supplementary Figure S13) and optical gap frequency in resonance with the LSP of LNP, which are ideal for strong coupling. We found that the $MoSe_2$-LNP hybrid exhibits a large (~25 %) transparency window in the vicinity of the A exciton peak as shown in Figure 5, as well as the $WSe_2$-LNP hybrid shown in Figure S14. The extinction features with a large transparency dip are similar to those previously observed in molecules hybridized with plasmonic nanostructures, where the features were attributed to Rabi splitting and emergence of plexcitons.[5,9,10] Strictly speaking, such hybridization modes (plexciton) appear when $\kappa > \Gamma_{LSP}, \Gamma_{exciton}$ is satisfied, in the strong coupling regime.[1,5,10] While we do not observe a clear evidence of mode splitting, the strong suppression of the extinction spectrum reveals effective plexcitonic coupling in this system.

In summary, we showed that exciton-plasmon coupling can be tuned in monolayer $MX_2$-Ag NP hybrid systems. Significant PL enhancement was achieved in monolayer $MoS_2$ by tuning LSPR to excitation and exciton energy. The low quantum yield of monolayer $MoS_2$ allows enhancement effects to overcome quenching even when the metal particles are in direct contact with the emitter. The PL enhancement in monolayer $WS_2$, which tends to exhibit high PL quantum yield, was only achieved when using a dielectric spacer layer to minimize quenching by the metal. Furthermore, exciton-induced transparency dips were observed due to the coherent dipole-dipole coupling between excitons of monolayer $MoS_2$ and LSP in resonance condition, as well as monolayer $MoSe_2$ and $WSe_2$. Our results highlight that integration of 2D semiconductors with plasmonic nanostructures is a viable approach to engineering light-matter interaction in these materials to enable their implementations in photonics. Tunable coupling between excitons and LSP from weak to strong coupling regimes is of fundamental interest and opens up prospects for novel technological applications.

*Experimental*

Experimental and FDTD simulation details are presented in the Supporting Information.

*Acknowledgements*

G.E acknowledges Singapore National Research Foundation for funding the research under NRF Research Fellowship (NRF-NRFF2011-02) and medium-sized centre programm. K. O. acknowledges support from JSPS KAKENHI (26289109). KM is supported by a Grant-in-Aid for Scientific Research from the Japan Society for Promotion of Science (Grants Nos. 26107522, 15K13500, 25400324, 25246010).

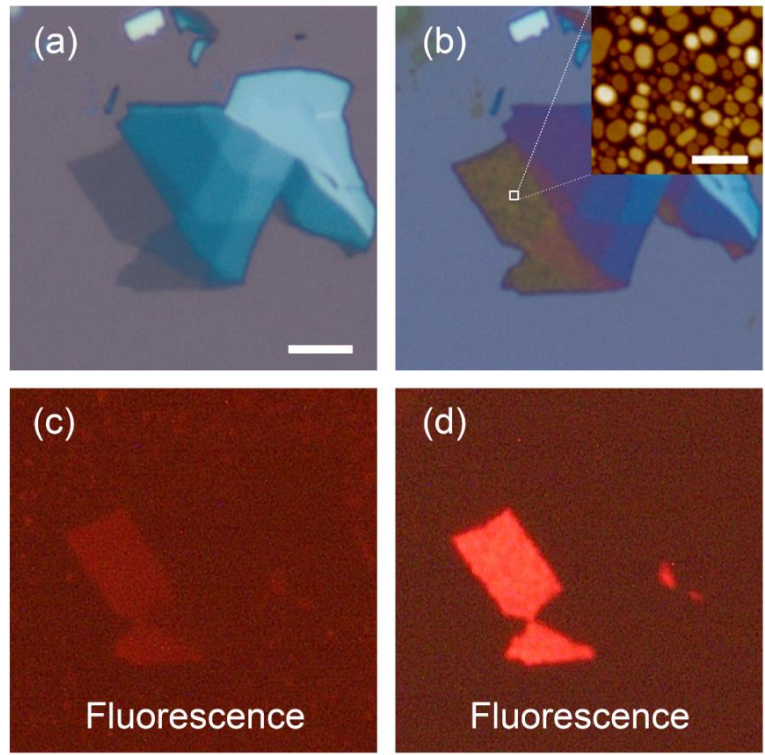

**Figure 1.** Figure 1 Optical images of bare monolayer $MoS_2$ (a) and $MoS_2$ with Ag nanoparticles on $SiO_2$(90nm)/Si wafer (b). Fluorescence images of bare monolayer $MoS_2$ (c) and $MoS_2$ with Ag nanoparticles (d). Due to low PL intensity of monolayer $MoS_2$, long exposure time (5s) and large gain were used to get clear fluorescence image for comparison. The insert image in (b) shows AFM image of Ag nanoparticles at the region indicated by the white square. The scale bar in (a) and insert of (b) are 4μm and 200nm, respectively.

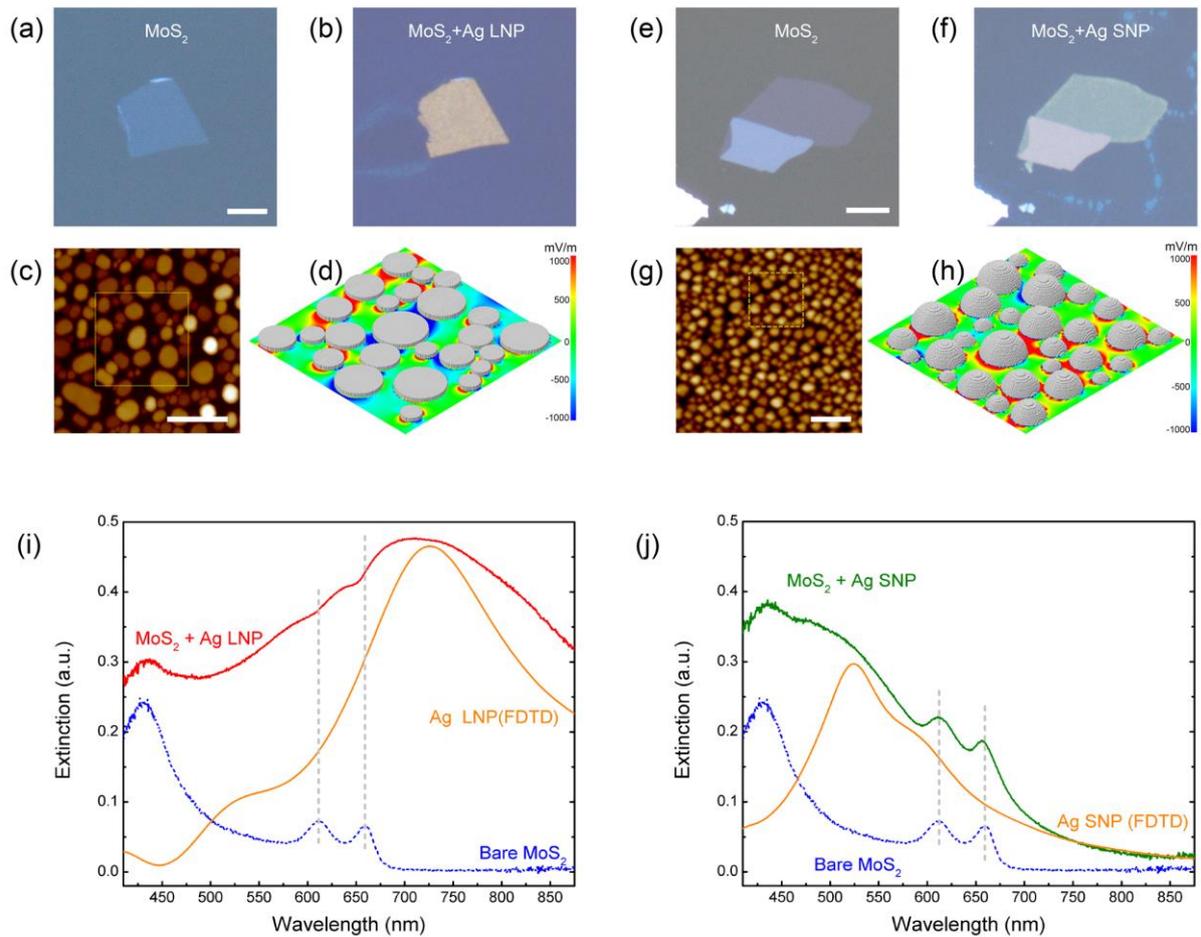

**Figure 2.** Optical images of monolayer $MoS_2$ on quartz without (a and e) and with (b and f) silver nanoparticles. AFM images (c and g) of monolayer $MoS_2$ with Ag nanoparticles and FDTD simulation results (d and h) for local electromagnetic field enhancement based on the simulated area indicated in the AFM images. The Ag LNP are simplified as nano- discs with 10nm thickness and varying diameter in our FDTD simulation, while Ag SNP are treated with semi-spheres with varying diameter, according to the AFM data. The maximum field enhancements are 52 and 40 times for the Ag LNP and SNP, respectively. The scale bars in optical and AFM images are 5μm and 200nm, respectively. (i) and (j) Experimental (red lines) and FDTD (orange lines) simulated extinction spectra of $MoS_2$ with Ag LNP and SNP, respectively, and experimental extinction of bare $MoS_2$ was shown as references (dashed blue line). The FDTD simulated extinction spectra were normalized with experimental results for comparison. The vertical dashed lines are guides for eyes.

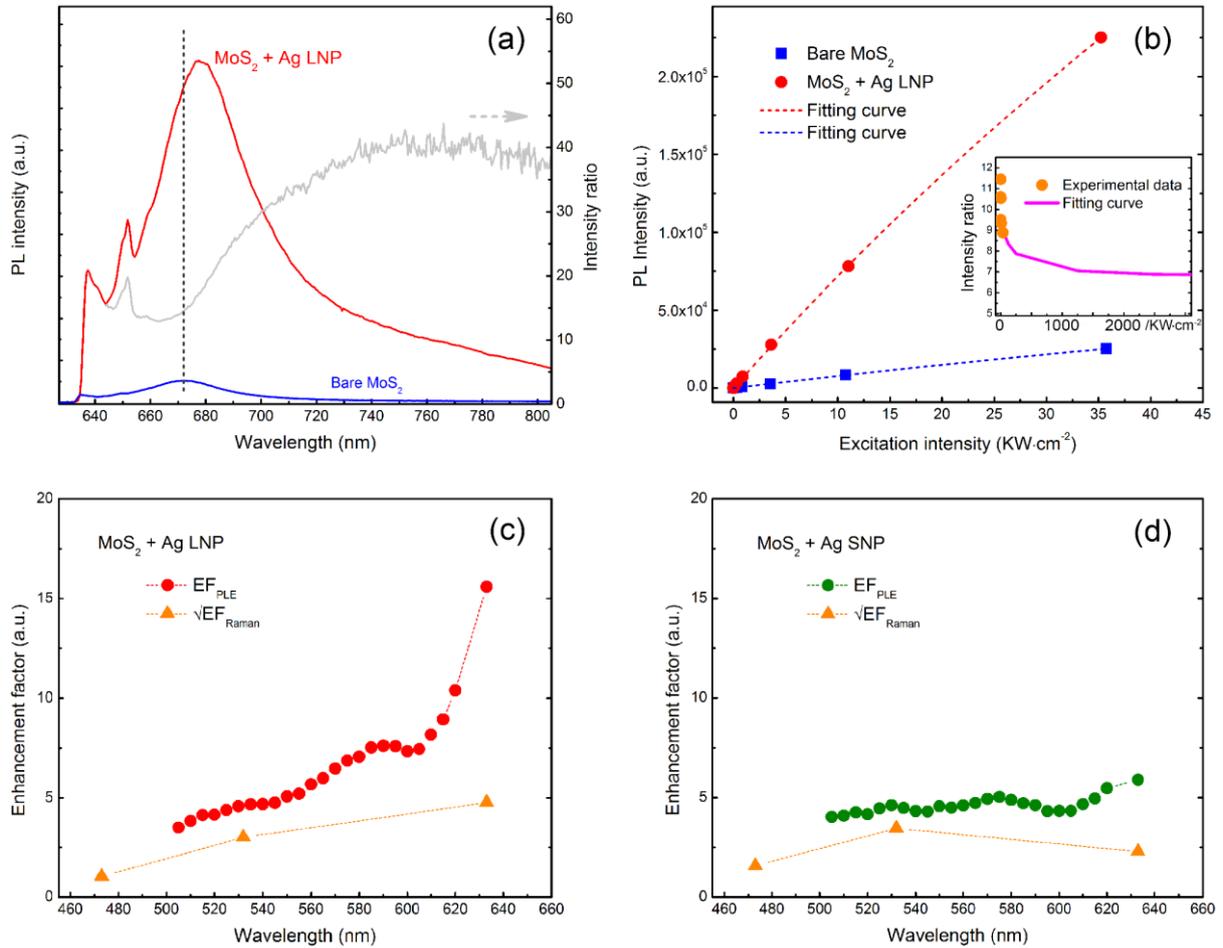

**Figure 3.** (a) PL spectra of monolayer $MoS_2$ with and without Ag SNP with 633nm laser excitation. The grey line is PL intensity enhancement ratio obtained from PL peak intensity of $MoS_2$ with Ag nanoparticles divided by that of bare $MoS_2$. The vertical dashed line is a guide for eyes. (b) Excitation power dependent PL of monolayer $MoS_2$ with and without Ag LNP at 633nm excitation. The experimental results (blue squares and red dots) are fitted with saturation curves (y=a*x/(b+x), where a and b are constants) shown as dotted lines. The insert image compares the power dependent PL enhancement factors obtained from experimental results (orange dots) and fitted curves (pink line). (c) and (d) Wavelength dependent PL enhancement factor of $MoS_2$ with Ag LNP and SNP, respectively, obtained from comparing photoluminescence excitation (PLE) spectra for $MoS_2$ with and without Ag nanoparticles.

Correspondingly, the square root of experimental Raman enhancement factor are shown for comparison.

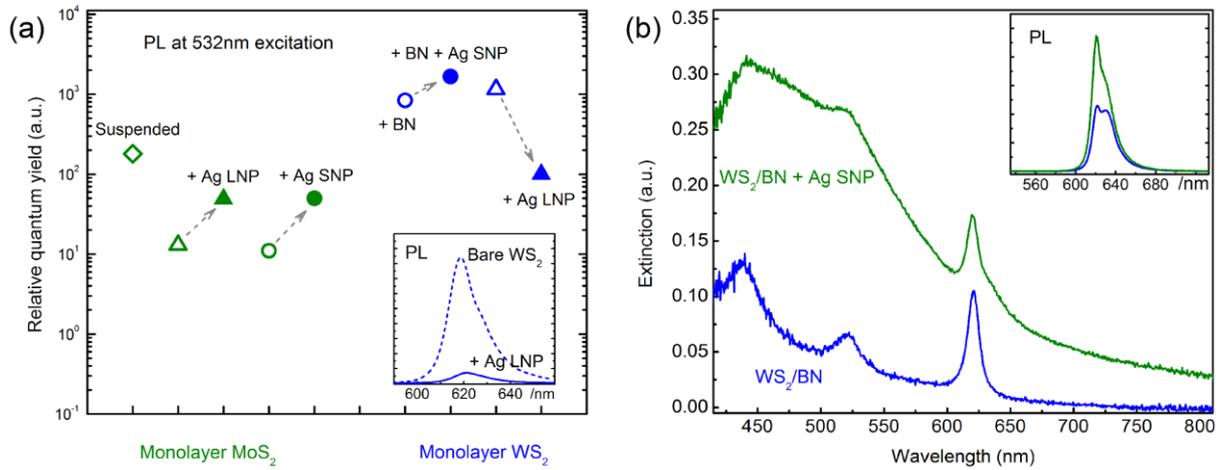

**Figure 4.** (a) Relative quantum yield of monolayer $MoS_2$ (green triangles and dots) and $WS_2$ (blue triangles and dots) on quartz substrate with and without Ag nanoparticles, as well as one suspended monolayer $MoS_2$ (green quadangles), at 532nm laser excitation, which are calculated by unit intensity of PL peak (counts·s$^{-1}$·μW$^{-1}$) divided by sample absorbance at 532nm. The absorbance at 532nm are ~5.4% and 4.5% for $MoS_2$ and $WS_2$ with h-BN, respectively, extracted from Figure 2i and 4b. The gray arrows indicate samples with and without silver nanoparticles. The insert image shows the PL quenching of monolayer $WS_2$ after Ag LNP deposition. (b) Extinction spectra of a heterostructure of monolayer $WS_2$ and h-BN film(~ 15nm thick) with (the olive line) and without (the blue line) Ag SNP, the insert image shows corresponding PL spectra. The optical and AFM images are shown in supporting information.

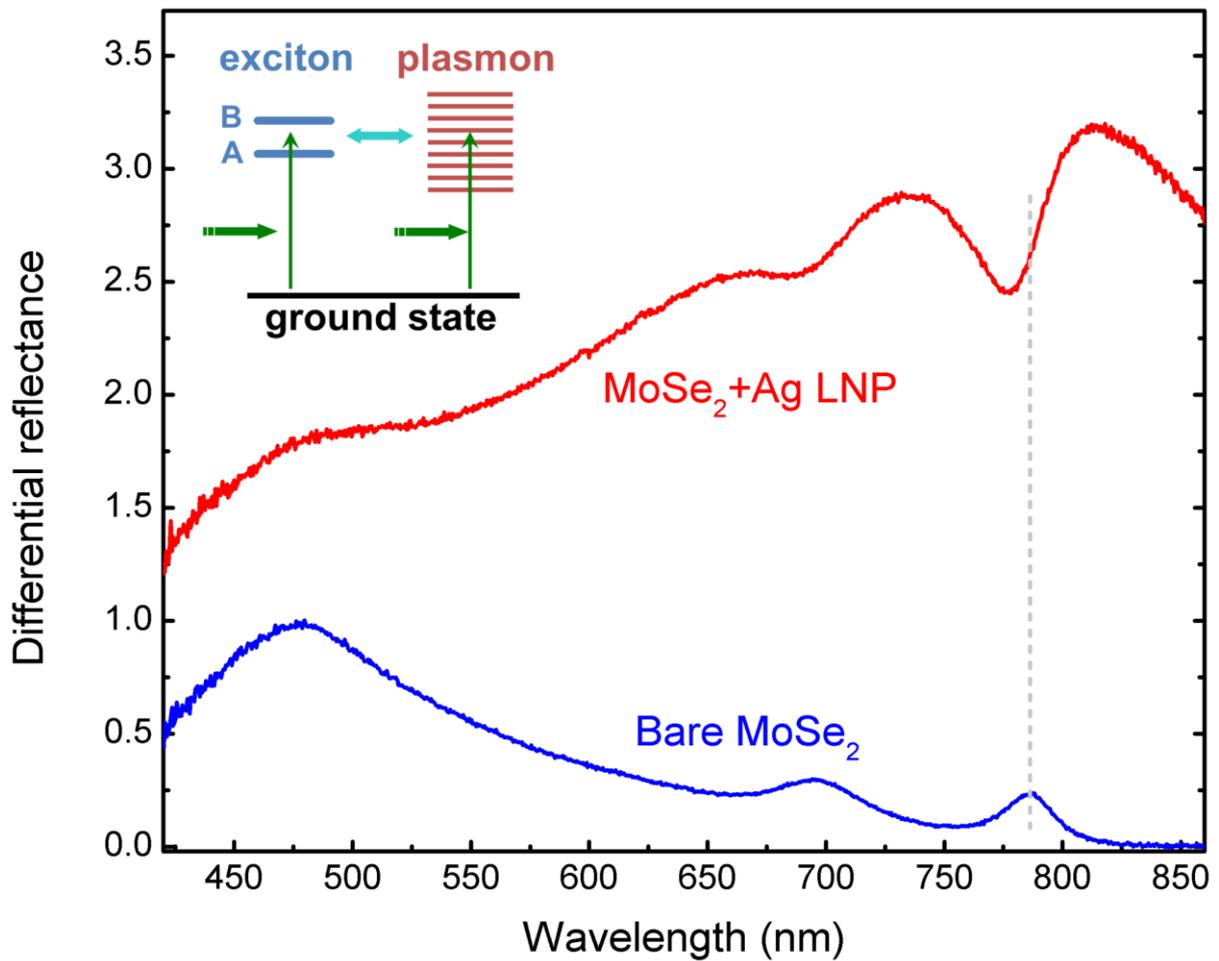

**Figure 5.** Differential reflectance (ΔR/R) of monolayer MoSe$_2$ with and without Ag LNP. The optical and AFM image of MoSe$_2$ are shown in supporting information. The dashed grey line is a guide for eyes. The insert image demonstrates the coherent dipole-dipole coupling between exciton of monolayer MoSe$_2$ and localized surface plasmon of silver nanoparticles.